\documentclass[11pt]{article}
\usepackage{graphicx}

\newcommand{\BABARPubYear}    {00}

\newcommand{\BABARConfNumber} {14}
\newcommand{\SLACPubNumber} {8536}

\usepackage{relsize}
\def\babar{\mbox{\slshape B\kern-0.1em{\smaller A}\kern-0.1em
    B\kern-0.1em{\smaller A\kern-0.2em R}}}

\def\en         {\ensuremath{e^-}}      
\def\ep         {\ensuremath{e^+}}
  
\def\epem       {\ensuremath{e^+e^-}}

\def\mumu       {\ensuremath{\mu^+\mu^-}}

\def\qqbar {\ensuremath{q\overline q}}

\def\piz   {\ensuremath{\pi^0}}

\def\pip   {\ensuremath{\pi^+}}
\def\pim   {\ensuremath{\pi^-}}
\def\pipi  {\ensuremath{\pi^+\pi^-}}

\def\Kbar  {\kern 0.2em\overline{\kern -0.2em K}{}}

\def\Kp    {\ensuremath{K^+}}
\def\Km    {\ensuremath{K^-}}

\def\Kzb   {\ensuremath{\Kbar^0}}
\def\KzKzb {\ensuremath{K^0 \kern -0.16em \Kzb}}
\def\Dz    {\ensuremath{D^0}}
\def\Dbar  {\kern 0.2em\overline{\kern -0.2em D}{}}

\def\Dzb   {\ensuremath{\Dbar^0}}
\def\DzDzb {\ensuremath{D^0 {\kern -0.16em \Dzb}}}

\def\Bz    {\ensuremath{B^0}}
\def\B     {\ensuremath{B}}
\def\Bbar  {\kern 0.18em\overline{\kern -0.18em B}{}}

\def\Bzb   {\ensuremath{\Bbar^0}}
\def\Bu    {\ensuremath{B^+}}
\def\Bub   {\ensuremath{B^-}}

\def\BB    {\ensuremath{B\Bbar}} 
\def\BzBzb {\ensuremath{B^0 {\kern -0.16em \Bzb}}}

\mathchardef\Upsilon="7107
\def\Y#1S{\ensuremath{\Upsilon{(#1S)}}}

\mathchardef\Deltares="7101
\mathchardef\Xi="7104
\mathchardef\Lambda="7103
\mathchardef\Sigma="7106
\mathchardef\Omega="710A
\def\Deltabar   {\kern 0.25em\overline{\kern -0.25em \Deltares}{}}
\def\Lbar {\kern 0.2em\overline{\kern -0.2em\Lambda\kern 0.05em}\kern-0.05em{}}
\def\Sigbar{\kern 0.2em\overline{\kern -0.2em \Sigma}{}}
\def\Xibar{\kern 0.2em\overline{\kern -0.2em \Xi}{}}
\def\Obar{\kern 0.2em\overline{\kern -0.2em \Omega}{}}
\def\Nbar{\kern 0.2em\overline{\kern -0.2em N}{}}
\def\Xbar{\kern 0.2em\overline{\kern -0.2em X}{}}

\def\BR{{\ensuremath{\cal B}}}

\def\mes        {\mbox{$m_{\rm ES}$}}

\def\ev   {\ensuremath{\rm \,e\kern -0.08em V}}
\def\kev  {\ensuremath{\rm \,ke\kern -0.08em V}} 
\def\mev  {\ensuremath{\rm \,Me\kern -0.08em V}} 
\def\gev  {\ensuremath{\rm \,Ge\kern -0.08em V}} 
\def\gevc {\ensuremath{{\rm \,Ge\kern -0.08em V\!/}c}} 
\def\tev  {\ensuremath{\rm \,Te\kern -0.08em V}}
\def\mevc {\ensuremath{{\rm \,Me\kern -0.08em V\!/}c}} 
\def\gevcc{\ensuremath{{\rm \,Ge\kern -0.08em V\!/}c^2}} 
\def\mevcc{\ensuremath{{\rm \,Me\kern -0.08em V\!/}c^2}}

\def\invfb   {\ensuremath{\mbox{\,fb}^{-1}}}
\def\mus  {\ensuremath{\rm \,\mus}}

\def\mus        {\ensuremath{\,\mu{\rm s}}}    
\def\gsim{{~\raise.15em\hbox{$>$}\kern-.85em
          \lower.35em\hbox{$\sim$}~}}
\def\lsim{{~\raise.15em\hbox{$<$}\kern-.85em
          \lower.35em\hbox{$\sim$}~}}

\def\CP                 {\ensuremath{C\!P}}

\def\to                 {\ensuremath{\rightarrow}}

\def\pep2{PEP-II}

\newcommand{\eqref}[1]{Eq.~(\ref{eq:#1})}

\newcommand{\npb}       [1]  {{Nucl.\ Phys.\ {\bf B{\bf #1}}}}

\newcommand{\prep}      [1]  {{Phys.\ Rep.{\bf #1}}}
\newcommand{\prl}       [1]  {{Phys.\ Rev.\ Lett.\ {\bf #1}}} 
\newcommand{\pr}        [1]  {{Phys.\ Rev.\ {\bf #1}}}

\newcommand{\zpc}       [1]  {{Z.\ Phys.\ C~{\bf #1}}}

\def\geant      {\mbox{\tt Geant321}}

\def\jetset74   {\mbox{\tt Jetset \hspace{-0.5em}7.\hspace{-0.2em}4}}

\def\Btohh     {\ensuremath{\B \to h^+h^-}}

\def\Bubtopipiz     {\ensuremath{\Bub \to \pim\piz}}
\def\Btopizpiz     {\ensuremath{\B \to \piz\piz}}

\def\pipi     {\ensuremath{\pi\pi}}
\def\kpi     {\ensuremath{K\pi}}
\def\kk     {\ensuremath{KK}}
\def\fish    {\ensuremath{\cal F}}

\def\cerenkov{$\check{\rm C}{\rm erenkov}$}

\long\def\inst#1{\par\nobreak\kern 4pt\nobreak
    {\it #1}\par\vskip 10pt plus 3pt minus 3pt}

\begin{document}
{\pagestyle{empty}

\begin{flushright}
\babar-CONF-\BABARPubYear/\BABARConfNumber \\
SLAC-PUB-\SLACPubNumber
\end{flushright}

\par\vskip 3cm

\begin{center}
\Large \bf Measurement of branching fractions for two-body charmless 
{\boldmath $B$} decays to charged pions and kaons at \babar\
\end{center}
\bigskip

\begin{center}
\large The \babar\ Collaboration\\
\mbox{ }\\
July 25, 2000
\end{center}
\bigskip \bigskip

\begin{center}
\large \bf Abstract
\end{center}
We present preliminary results of a search for charmless two-body \B\ decays
to charged pions and kaons using data collected by the \babar\ detector
at the Stanford Linear Accelerator Center's PEP-II storage ring.  In
a sample of 8.8 million produced $\BB$ pairs we measure the branching
fractions 
$\BR(\Bz\to \pip\pim) = (9.3^{+2.6}_{-2.3}$$^{+1.2}_{-1.4})\times 10^{-6}$ and
$\BR(\Bz\to \Kp\pim) = (12.5^{+3.0}_{-2.6}$$^{+1.3}_{-1.7})\times 10^{-6}$,
where the first uncertainty is statistical and the second is systematic. 
For the decay $\Bz\to\Kp\Km$ we find no significant signal and set an upper 
limit of $\BR(\Bz\to\Kp\Km) < 6.6 \times 10^{-6}$ at the $90\%$ confidence 
level.

\vfill
\begin{center}
Submitted to the XXX$^{th}$ International Conference on High Energy Physics, 
Osaka, Japan.
\end{center}

\newpage
}

\begin{center}
\small

The \babar\ Collaboration
\bigskip

B.~Aubert,
A.~Boucham,
D.~Boutigny,
I.~De Bonis,
J.~Favier,
J.-M.~Gaillard,
F.~Galeazzi,
A.~Jeremie,
Y.~Karyotakis,
J.~P.~Lees,
P.~Robbe,
V.~Tisserand,
K.~Zachariadou
\inst{Lab de Phys.\ des Particules, F-74941 Annecy-le-Vieux, CEDEX, France}
% ...

\end{center}\newpage

\setcounter{footnote}{0}

\section{Introduction}
\label{sec:Introduction}

Measurements of the branching fractions for the rare charmless decays 
$\Bz\to h^+h^-\,(h=\pi,K)$\footnote{Charge conjugate decay modes are 
assumed throughout this paper.} provide important information in the 
study of charge-parity (\CP) violation.  In principle, the \pip\pim\ decay 
mode can be used to extract the angle $\alpha$ of the Unitarity Triangle 
through the phenomenon of \Bz\--\Bzb\ mixing.  However, in addition to 
the dominant $b\to uW^-$ tree amplitude, this decay includes the 
$b\to dg$ penguin amplitude and the determination of $\alpha$ is 
subject to large theoretical uncertainties if the penguin contribution is 
non-negligible~\cite{ref:gronau1}.  In the presence of significant 
``penguin pollution,'' additional measurements of the isospin-related
decays \Bubtopipiz\ and \Btopizpiz\ provide a means of measuring 
$\alpha$ cleanly in the \pipi\ channel~\cite{ref:gronau2}.

The decay \Bz\to\Kp\pim\ is dominated by the $b\to sg$ penguin amplitude and 
provides an estimate of the scale of penguin pollution in the \pip\pim\ 
decay.  Recent results from the CLEO collaboration indicate that 
the decay rate for \Bz\to\Kp\pim\ is significantly larger than the
rate for \Bz\to\pip\pim\ \cite{ref:CLEO1}, implying that the
penguin contribution in the $\pi\pi$ decay is indeed significant and an 
isospin analysis will be necessary to measure $\alpha$ accurately.
The apparent enhancement of the penguin amplitude improves the 
prospects for observing direct \CP\ violation as an asymmetry in the 
decay rates for \Bz\to\Kp\pim\ and \Bzb\to\Km\pip.  
Precise measurement of the decay rates for \pipi\ and \kpi\ decays is 
therefore of central importance.  This paper describes preliminary 
measurements of branching fractions for the decays \Bz\to\pip\pim, 
\Kp\pim, and \Kp\Km\ with the first data collected by the \babar\ 
experiment.  

\section{Data sample, \babar\ detector, and event selection}
\label{sec:babar}

The dataset used in this analysis consists of $8.9\invfb$ collected with the 
\babar\ detector at the Stanford Linear Accelerator Center's PEP-II 
storage ring between January and June 2000.  The PEP-II
facility operates nominally at the \Y4S\ resonance, providing 
asymmetric collisions of $9.0\gev$ electrons on $3.1\gev$ positrons.  The 
dataset includes $7.7\invfb$ collected in this configuration (on-resonance) and 
$1.2\invfb$ collected below the \BB\ threshold (off-resonance) that are used for 
continuum background studies.  The on-resonance sample corresponds to 8.8 million
produced \BB\ events.  

The asymmetric beam configuration in the laboratory frame provides a boost 
$(\beta\gamma = 0.56)$ to the \Y4S, allowing separation of the $\B$ and 
$\Bbar$ decay 
products for time-dependent $\CP$-violation studies.  For the analysis 
described in this paper, the significant effect of the boost 
relative to symmetric collider experiments is to increase the momentum range 
of the \B\ decay products from a narrow distribution centered at 
$\sim 2.6\gevc$, to a broad distribution extending from $1.5$ to $4.5\gevc$.  
Wherever necessary, kinematic quantities evaluated in the \Y4S\ center-of-mass 
(CM) frame are denoted with an additional superscript asterisk in order to 
distinguish them from the corresponding quantities evaluated in the laboratory 
frame.

\babar\ is a solenoidal detector optimized for the asymmetric beam configuration
at PEP-II and is described in detail elsewhere~\cite{ref:babar}.  Charged 
particle (track) momenta are measured in a tracking system consisting of a 
5-layer, double-sided, silicon vertex tracker and a 40-layer drift chamber 
filled with a gas mixture of helium and isobutane, both operating within a 
$1.5\,{\rm T}$ superconducting solenoidal magnet.  Photons are detected in an 
electromagnetic calorimeter consisting of 6580 CsI(Tl) crystals arranged in 
barrel and forward endcap subdetectors that also operates within the magnetic 
field.  The iron used for the magnet yoke is segmented and instrumented with 
resistive plate chambers, providing muon identification and (in conjunction 
with the calorimeter) neutral hadron detection.  

In this analysis, tracks are identified as pions or kaons by the \cerenkov\ angle 
$\theta_c$ measured by a Detector of Internally Reflected \cerenkov\ light (DIRC).  
The DIRC system is a unique type of \cerenkov\ detector that relies on total internal 
reflection within the radiator to deliver the \cerenkov\ light outside the 
tracking and magnetic volumes.  The radiator consists of 144 axially aligned 
synthetic quartz bars located just inside the inner radius of the calorimeter.  
The bars extend outside the solenoid flux return in the backward direction, where 
the \cerenkov\ ring is imaged by an array of $\sim 11000$ photomultiplier tubes.  

Hadronic events are selected based on track multiplicity and event
topology.  Tracks with transverse momentum greater than $100\mevc$ are required 
to pass efficient quality cuts, including number of drift chamber hit layers 
used in the track fit and impact parameter in the $r$--$\phi$ and $r$--$z$ 
planes, where the cylindrical coordinate $z$ is aligned along the detector axis 
in the electron beam direction.  At least three tracks must pass the above 
selection.  To reduce contamination from Bhabha and \mumu\ events the ratio of 
second to zeroth Fox-Wolfram moments~\cite{ref:foxwolf}, $R_2 = H_2/H_0$, is 
required to be less than $0.95$.  Residual background from tau hadronic decays 
is reduced by requiring the sphericity~\cite{ref:sphericity} of the event to be greater 
than $0.01$.  The efficiency of the event selection is dominated by the
acceptance and efficiency of the track requirement, and is determined to be 
$70\%$ from a detailed Monte Carlo simulation based on \geant\ 
\cite{ref:geant}.

\section{\boldmath Selection of \Btohh candidates}
\label{sec:selection}

One of the advantages of measuring \B\ decay parameters at the \Y4S\ resonance
is the kinematic constraint provided by the initial state, where energy 
conservation determines that the energies of the \B\ mesons in the CM frame are 
equal to $\sqrt{s}/2$, where $\sqrt{s}$ is the total $\ep\en$ CM energy.  We 
exploit this constraint by calculating an energy-substituted mass \mes, where
$\sqrt{s}/2$ is substituted for the \B\ candidate energy, and by calculating 
the energy difference $\Delta E$ between the \B\ candidate and $\sqrt{s}/2$ in 
the CM frame.

We define $\mes = \sqrt{(\sqrt{s}/2)^2 - p_B^{*2}}$, where $p^*_B$ is the \B\ 
candidate momentum evaluated in the CM frame.  Because $p^*_B$ is relatively 
small $(\sim 300\mevc)$, the resolution on \mes\ is dominated by the 
uncertainty in $\sqrt{s}$, which in turn is determined by the beam energy 
spread and width of the \Y4S\ resonance.  Substitution of the beam energy 
reduces the mass resolution by one
order of magnitude compared to the invariant mass.  The mean value of \mes\ and 
its Gaussian width $\sigma(\mes)$ are determined from a large sample of fully 
reconstructed \B\ decays.  We find $\mes = (5.2800\pm 0.0005)\gevcc$ and 
$\sigma(\mes) = (2.6\pm 0.1)\mevcc$, respectively.  Our initial selection 
requires $5.22 < \mes < 5.30\gevcc$.

We define $\Delta E = E_B^* - \sqrt{s}/2$, where $E_B^*$ is the \B\
candidate energy in the CM frame.  Signal events are Gaussianly distributed in
$\Delta E$ with a mean near zero, while the continuum background falls linearly 
over the region of interest.  For this analysis, the pion mass is assigned to
all tracks and the $\kpi$ and $\kk$ decays have $\Delta E$ shifted from zero 
by an amount depending on the momentum of the tracks.  From Monte Carlo 
simulation we find shifts of $-45$ and $-91\mev$ for the $\kpi$ and $\kk$ 
decays, respectively.  The resolution on $\Delta E$ is estimated to be 
$27\pm 5\mev$ based on Monte Carlo simulated \pipi\ decays and the observed 
difference in widths between data and Monte Carlo \Bub\to\Dz\pim\ decays.  
We require $\left| \Delta E \right| < 0.420\gev$.

\section{Background suppression}
\label{sec:bkg}

Due to the relatively small CM momenta of decay products produced from the quark 
transition $b\to c$, \B\ decays to final states involving charm mesons are not
a significant background to charmless two-body decays.  After hadronic selection, 
the background is dominated by continuum production of light quarks, 
$\epem\to\qqbar\,(q = u, d, s, c)$.  In the CM frame, the continuum background 
typically exhibits a two-jet structure that can produce two high momentum 
back-to-back tracks with an invariant mass near the \B\ mass.  In contrast, the 
low momentum of \B\ mesons in the decay \Y4S\to\BB\ leads to 
a more spherically symmetric event.  
This topology difference is exploited by constructing the angle $\theta_S$ 
between the sphericity axes, evaluated in the CM frame, of the \B\ candidate 
and the remaining charged and neutral particles in the event.  The absolute 
value of the cosine of this angle is strongly peaked near 1 for continuum 
events and is approximately uniform for \BB\ events.  We require 
$\left| \cos{\theta_S} \right| < 0.9$, which is $87\%$ efficient for signal 
events and rejects $66\%$ of the continuum background.  

Further separation power between signal and continuum background is provided by a
Fisher discriminant technique~\cite{ref:CLEO2}.  The Fisher discriminant \fish\ is 
calculated from a linear combination of discriminating variables
$x_i$,
\begin{equation}
{\cal F} = \sum_{i=1}^{9} \alpha_i x_i,
\end{equation}
where the coefficients $\alpha_i$ are chosen to maximize the statistical 
separation between signal and background events.  The nine discriminating 
variables are constructed from the scalar sum of the momenta of all charged 
and neutral particles (excluding the candidate daughter tracks) flowing 
into nine concentric cones centered on the $B$-candidate thrust axis in the CM
frame.  Each cone subtends an angle of $10^{\circ}$ and is 
folded to combine the forward and backward intervals.  More energy will be 
found in the cones nearer the candidate thrust axis in jet-like continuum 
background events than in the more isotropic \BB\ events.

\begin{figure}[!htb]
\begin{center}
\includegraphics[height=8cm]{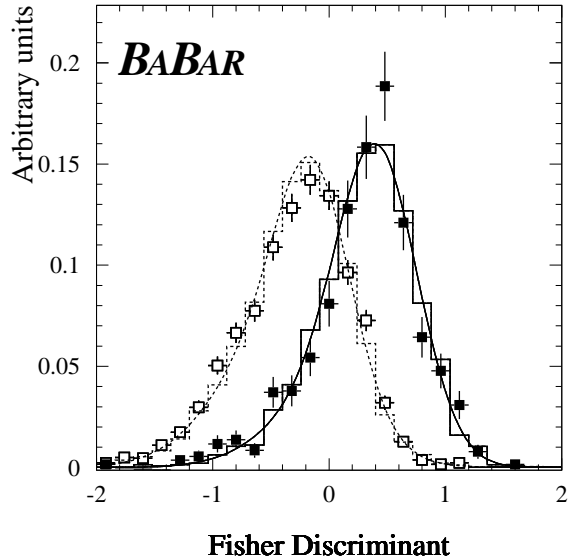}
\caption{The \fish\ distribution for $\pipi$ signal Monte Carlo events (solid
histogram and fitted curve) compared to data $\Bub\to \Dz\pim$ decays 
(filled squares), and continuum background Monte Carlo (dashed histogram and
fitted curve) compared to off-resonance data (open squares).  The Monte Carlo
samples are independent of the samples used to train the Fisher 
discriminant.}
\label{fig:fisher}
\end{center}
\end{figure}

Large samples of signal and background Monte Carlo simulated events 
reconstructed in the \pipi\ mode are used to determine the Fisher coefficients.  
Figure~\ref{fig:fisher} shows the resulting \fish\ distributions for signal 
\pipi\ Monte Carlo compared to a sample of \Bub\to\Dz\pim\ decays reconstructed 
in data, and continuum background Monte Carlo compared to off-resonance data.  
The \fish\ distributions for both signal and background are parameterized
by the sum of two Gaussians with separate means and widths.  The Gaussian
fits are performed on Monte Carlo samples that are independent of the
samples used to determine the Fisher coefficients, and the same parameterization
is used for all three signal modes.  

\section{Particle identification}
\label{sec:pid}

Two complementary methods of exploiting the particle identification 
capabilities of the DIRC are described in this paper.  The first method 
uses measurements of $\theta_c$ to derive particle {\em selectors} that are used 
to identify pions and kaons on a per-track basis.  The second method uses 
likelihood functions derived from $\theta_c$ measurements directly in a 
maximum-likelihood fit to extract the relative amount of each decay 
mode on a statistical basis.  After the selection criteria described above, the 
combined acceptance and efficiency of requiring a $\theta_c$ measurement for 
both tracks is $76\%$.

A control sample consisting of $18141\pm 140$ \Dz\to\Km\pip\ candidate decays 
is used to parameterize and assess the performance of both particle 
identification methods.  A $96\%$ pure \Dz\ sample is obtained through the 
decay $D^{*+}\to \Dz\pip_s$, 
where the slow pion $\pip_s$ tags the charge of the pion from the \Dz\ decay.  
Requiring a tight window around the $D^{*+}$--\Dz\ mass difference minimizes 
contamination from incorrectly reconstructed \Dz\ mesons, where the two daughter 
tracks are assigned the incorrect particle hypotheses.  The momentum of kaons
and pions in the control sample ranges from $1.75$--$4.0\gevc$, which covers 
$90\%$ of the momentum range for charmless two-body \B\ decays.

\begin{figure}[!htb]
\begin{center}
\includegraphics[height=8cm]{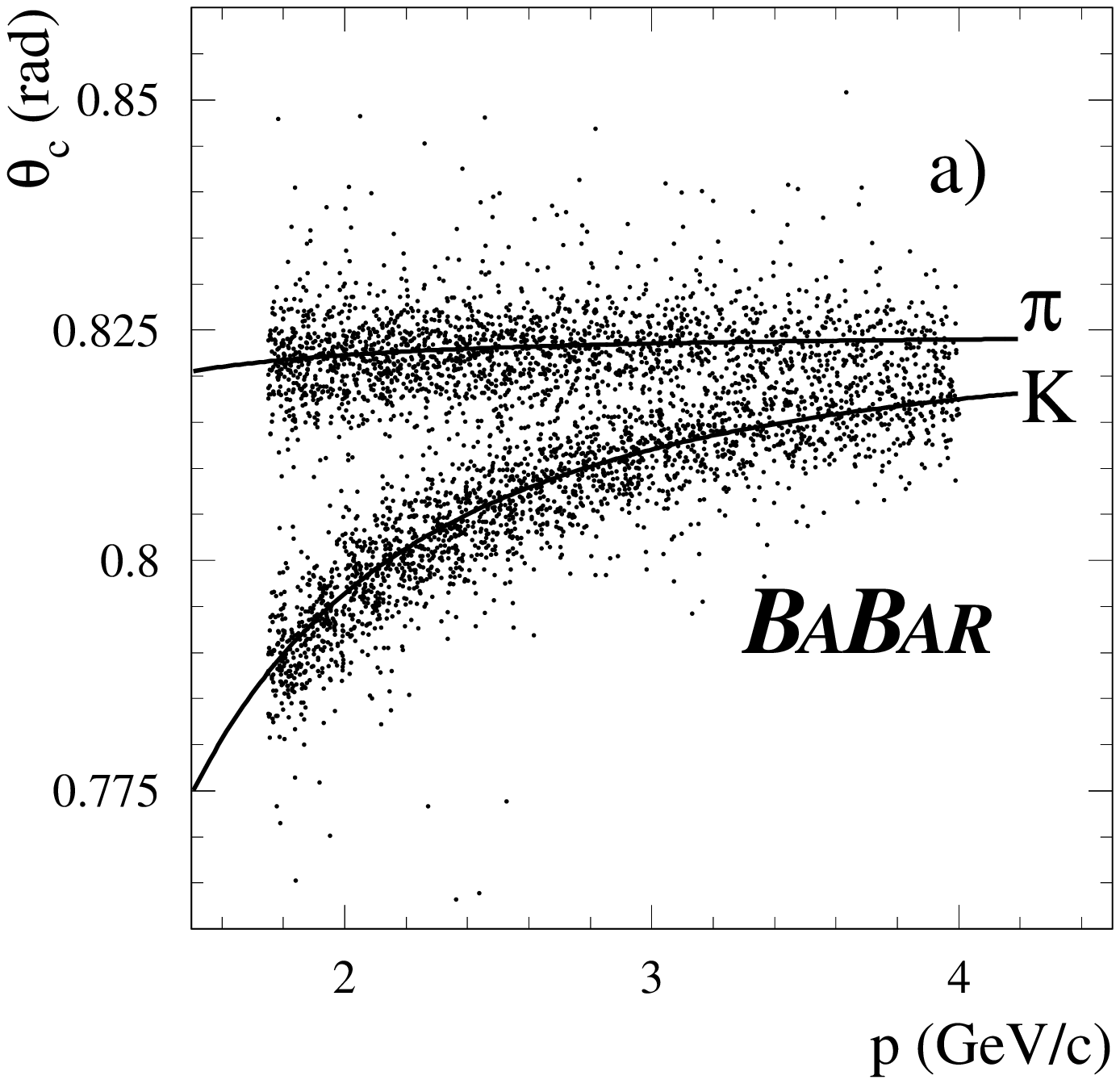}
\includegraphics[height=8cm]{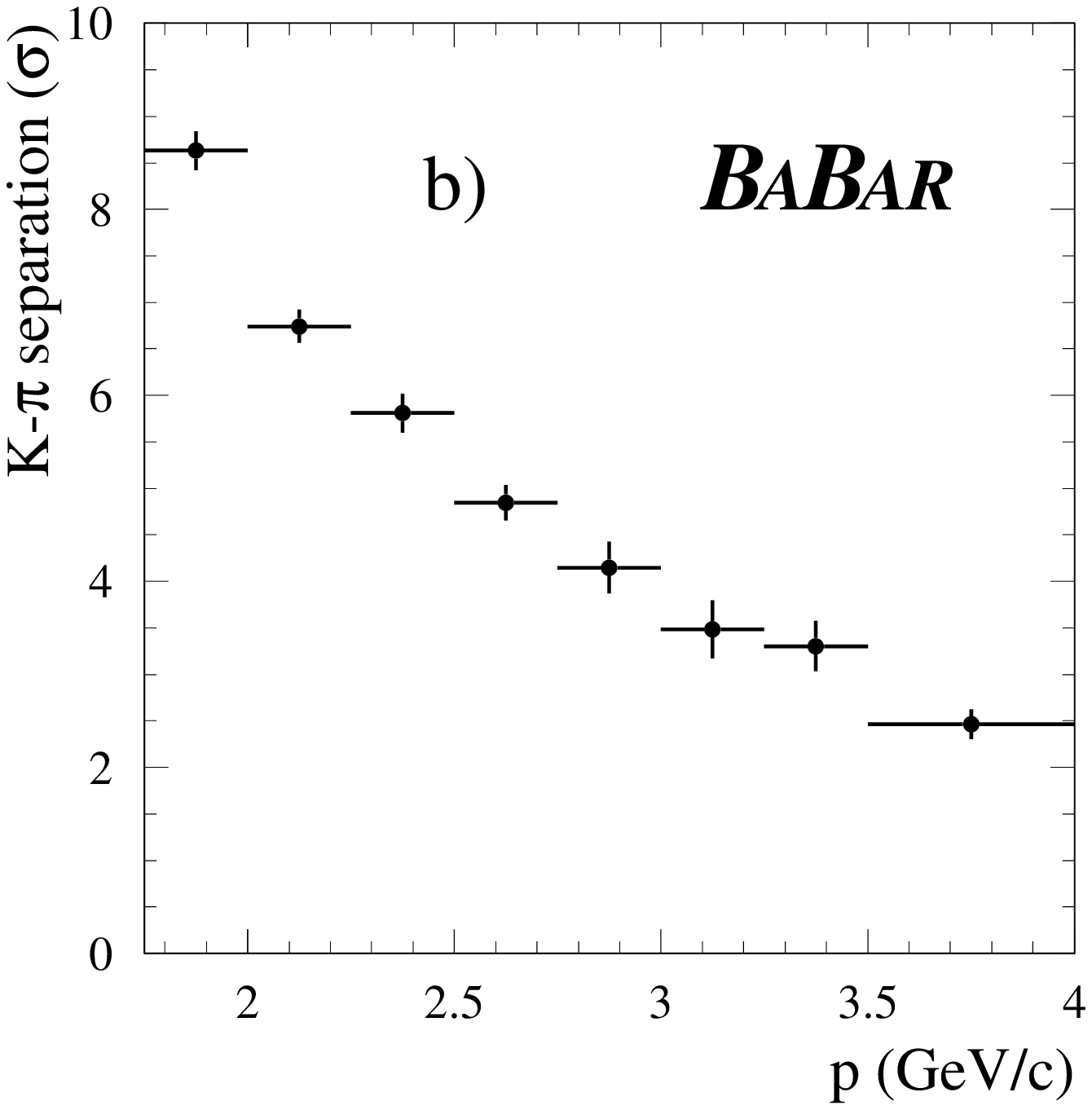}
\caption{(a) The \cerenkov\ angle and (b) $K$--$\pi$ separation as
functions of momentum for single tracks in the \Dz\ control sample.
The separation is an average over all polar angles.}
\label{fig:dirc}
\end{center}
\end{figure}

The performance of the DIRC is summarized in Fig.~\ref{fig:dirc}, where we show 
plots of (a) $\theta_c$ vs. $p$ for tracks in the control sample and (b) the
statistical separation between pions and kaons as a function of momentum.  
The separation, defined as 
$\left[\langle \theta_c(\pi)\rangle - \langle \theta_c(K)\rangle \right]
/\langle \sigma_{\theta_c}\rangle$, where $\langle\theta_c(h)\rangle$ are the
Gaussian means for pion and kaon tracks and $\langle\sigma_{\theta_c}\rangle$ is
the average width, varies from 8 at $2\gevc$ to $2.5$ at $4\gevc$.  Note
that the quoted $\theta_c$ separation is for a single track and an average 
over the polar angle of each track is implicit in Fig.~\ref{fig:dirc}(b).

The selector method of particle identification attempts to identify pions
and kaons on a per-track basis by cutting on likelihood functions.  We 
construct the likelihood for a given mass hypothesis from the 
Gaussianly-distributed $\theta_c$ measurements and the Poissonian probability for 
the number of measured \cerenkov\ photons compared to its expectation value.  
Tracks with momentum below the kaon threshold or within $2\sigma(\theta_c)$ of 
the expected value for a proton are explicitly removed, where the measurement 
error $\sigma(\theta_c)$ varies with momentum and the number of photons used in 
the fit.  We calculate the efficiency for one mode to be identified as another 
using the efficiency and mis-identification probabilities for pions and kaons 
measured in the \Dz\ control sample.  The resulting efficiency matrix is shown 
in Table~\ref{tab:effmisid}.  The matrix includes the probability that a pion 
or kaon is mis-measured and removed by the proton rejection requirement.

\begin{table}[!htb]
\caption{Particle selector efficiency and mis-identification probability for
each decay mode.  Errors are statistical only.}
\begin{center}
\smallskip
\begin{tabular}{cccc} \hline\hline
 & \multicolumn{3}{c}{Probability to be identified as} \\
 Mode & $\pipi$ & $\kpi$ & $\kk$ \\\hline
 $\pipi$ & $0.853 \pm 0.005$ & $0.109 \pm 0.008$ & $0.0029\pm 0.0002$ \\
 $\kpi$  & $0.121\pm 0.004$ & $0.775\pm 0.006$ & $0.051\pm 0.004$ \\
 $\kk$   & $0.0112\pm 0.0006$ & $0.231 \pm 0.008$ & $0.704\pm 0.007$ \\
 \hline\hline
\end{tabular}
\end{center}
\label{tab:effmisid}
\end{table}

The global likelihood method incorporates the particle identification 
probabilities for pions and kaons directly in an unbinned maximum likelihood fit.  
The probability density functions are constructed from $\theta_c$ 
information alone, where Gaussian fits are performed to the distribution of
${\rm measured} - {\rm expected}$ \cerenkov\ angle in bins of momentum in the \Dz\ 
control sample.  To improve the resolution and minimize the non-Gaussian tails of
the $\theta_c$ distribution, we require a minimum number of observed \cerenkov\ 
photons above the expected background.  From the control sample we determine the 
efficiency of this requirement to be $77\%$, taking into account the angular 
correlation between the two daughter tracks from a \B\ decay.  The proton 
rejection cut applied in the selector method is also required for the 
likelihood analysis.  After the minimum photon cut, the proton requirement is 
$98\%$ efficient for $h^+h^-$ events.

\section{Analysis}
\label{sec:analysis}

In this section we present the results of a simple cut-based analysis to determine
the $h^+h^-$ yield, followed by two complementary methods for determining the 
$\pipi$, $\kpi$, and $\kk$ yields in our data.  In the first method we 
apply background suppression and particle identification cuts to isolate
samples of events that are consistent with the $\pipi$, $\kpi$, or $\kk$ 
hypotheses.  Signal yields are then obtained from an unbinned maximum 
likelihood fit to \mes.  In the second method we perform a global maximum 
likelihood fit incorporating $\mes$, $\Delta E$, and ${\cal F}$, as well as the 
$\theta_c$ probability density functions described in the previous section, to 
determine signal yields in all three modes simultaneously.  The cut-based method 
is more transparent while the global fit has higher efficiency and statistical 
significance.  In Sec.~\ref{sec:br} we calculate branching fractions using the 
global fit results.

\subsection{Cut-based analysis}
\label{sec:cut}

In addition to the selection criteria described in Secs.~\ref{sec:selection} and
\ref{sec:bkg}, we tighten the $\theta_S$ requirement, 
$\left| \cos{\theta_S}\right| <0.7$, 
and also require $\left| \cos{\theta_B}\right|<0.8$, where $\theta_B$ 
is the CM angle of the \B\ candidate with respect to the beam.  The variable 
$\cos{\theta_B}$ is uniform for the continuum background and follows a 
$1-\cos^2{\theta_B}$ distribution for signal \B\ decays.  We determine an 
optimal selection requirement of ${\cal F} > 0.37$ by maximizing the statistical significance of the 
expected signal yield in a sample of Monte Carlo simulated signal and 
background events.  The relative efficiency of these additional cuts is $44\%$.  
Within the $\Delta E$--$\mes$ plane we define a signal region 
$\left| \Delta E \right| < 0.140\gev$ and $\mes = 5.2800 \pm 0.0052\gevcc$, 
and sideband regions $\left| \Delta E \right| > 0.140\gev$ and 
$\mes < 5.27\gevcc$.  The signal region in $\Delta E$ is designed to minimize 
contamination from three-body charmless \B\ decays where one of the decay
products has low momentum.

\begin{figure}[!htb]
\begin{center}
\includegraphics[height=6.cm]{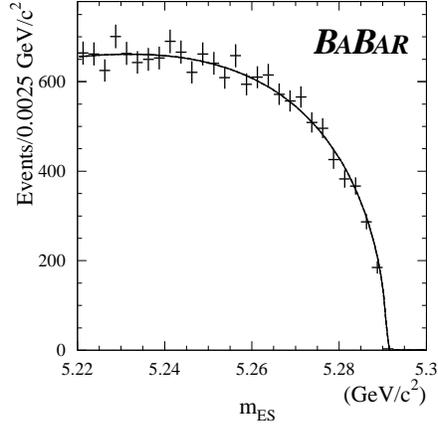}
\caption{Result of fitting of Eq.~\ref{arguseq} to the $\Delta E$ 
sideband region in on-resonance data.  The result is $\xi = 22.0\pm 0.5$.}
\label{fig:argusfit}
\end{center}
\end{figure}

\begin{figure}[!htb]
\begin{center}
\includegraphics[height=6.cm]{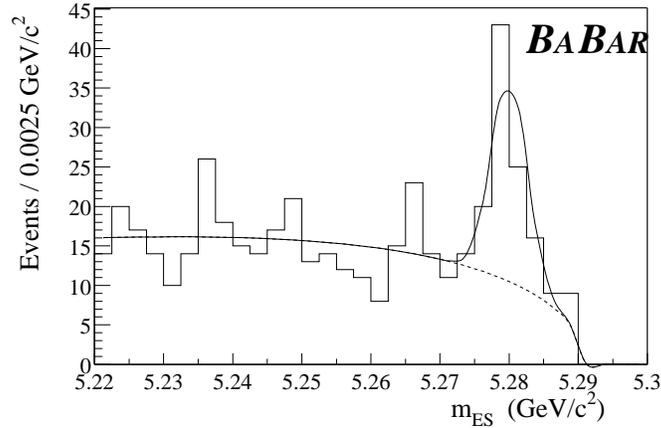}
\caption{Result of fitting the $\mes$ signal region in the $h^+h^-$ sample
to the sum of Eq.~\ref{arguseq}, with $\xi = 22$, and a Gaussian signal peak 
with $\mes = 5.28\gevcc$ and $\sigma(\mes) = 2.6\mevcc$.  The dashed curve 
shows the background parameterization only.}
\label{fig:hh}
\end{center}
\end{figure}

Before applying the particle selector we first demonstrate the presence
of $h^+h^-$ decays in the signal region.  The signal yield is obtained from 
an unbinned maximum likelihood fit to the $\mes$ distribution.  The fit 
includes candidates passing all cuts except the requirement that the tracks \
have an associated $\theta_c$ measurement.  The background shape in $\mes$ is 
parameterized by the empirical formula~\cite{ref:argus},
\begin{equation}
f(\mes) \propto \mes \sqrt{1-x^2} \exp{\left[ -\xi (1-x^2)\right]},
\label{arguseq}
\end{equation}
where $x = 2\mes/\sqrt{s}$ and the parameter $\xi$ is determined from a fit.
Figure~\ref{fig:argusfit} shows the results of fitting Eq.~\ref{arguseq} to the
on-resonance $\Delta E$ sideband region, where we find $\xi = 22.0\pm 0.5$.  We 
then fit the $\mes$ distribution in the $\Delta E$ signal region to a Gaussian
with mean and width fixed to $5.280\gevcc$ and $2.6\mevcc$, respectively, and
the background shape fixed to $\xi = 22$; only the normalizations are free parameters.  
The result is shown in Fig.~\ref{fig:hh}.  The fitted number of $h^+h^-$ candidates 
is $67\pm 11$, where the error is the statistical uncertainty from the fit.
Correcting for the total efficiency of the selection criteria ($27\%$) and 
normalizing to the total number of $\BB$ pairs, we determine a 
branching fraction $\BR(\Bz\to h^+h^-) = (28 \pm 5)\times 10^{-6}$, where 
the error is statistical only and we assume equal \Y4S\ branching fractions to 
charged and neutral \B\ mesons.

\begin{figure}[!htbp]
\begin{center}
\includegraphics[height=7cm]{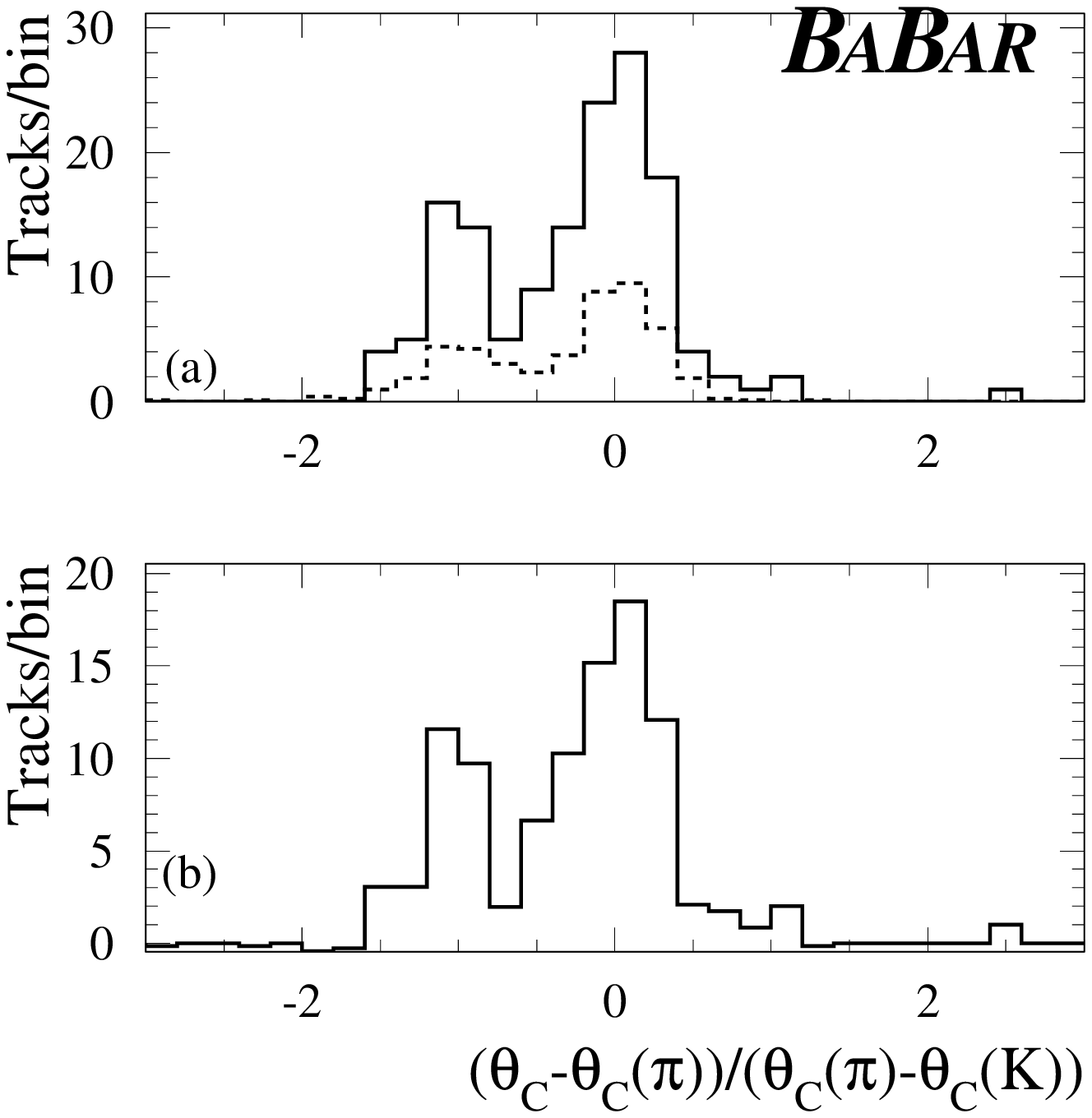}
\caption{The measured \cerenkov\ angle minus its expectation value for the pion
hypothesis, normalized to the expected separation between pions and kaons.  
True pions (kaons) appear as a peak at $0\,(-1)$.  (a) shows the 
distribution for \B\-candidate tracks in the signal region of $\Delta E$ and
the signal (solid) and sideband (dashed) regions of $\mes$, while (b) 
shows the signal region after sideband subtraction.  Note that both tracks 
from each \B\ candidate are included in these plots.}
\label{fig:kaonproof}
\end{center}
\end{figure}

The presence of kaons in the signal region can be demonstrated by plotting the 
variable $\left( \theta_c - \theta_c(\pi)\right)/\left(\theta_c(\pi) - \theta_c(K)\right)$,
where $\theta_c(\pi)$ and $\theta_c(K)$ are the expected values.  This variable
peaks at $0\,(-1)$ for real pion\,(kaon) tracks.  In Fig.~\ref{fig:kaonproof} we 
show the distribution of this variable for $B$-candidate tracks in the $\mes$ 
signal region before and after subtracting the distribution obtained from
the $\mes$ sideband region.  A clear kaon peak remains after subtraction.

\begin{figure}[!htbp]
\begin{center}
\includegraphics[height=9cm]{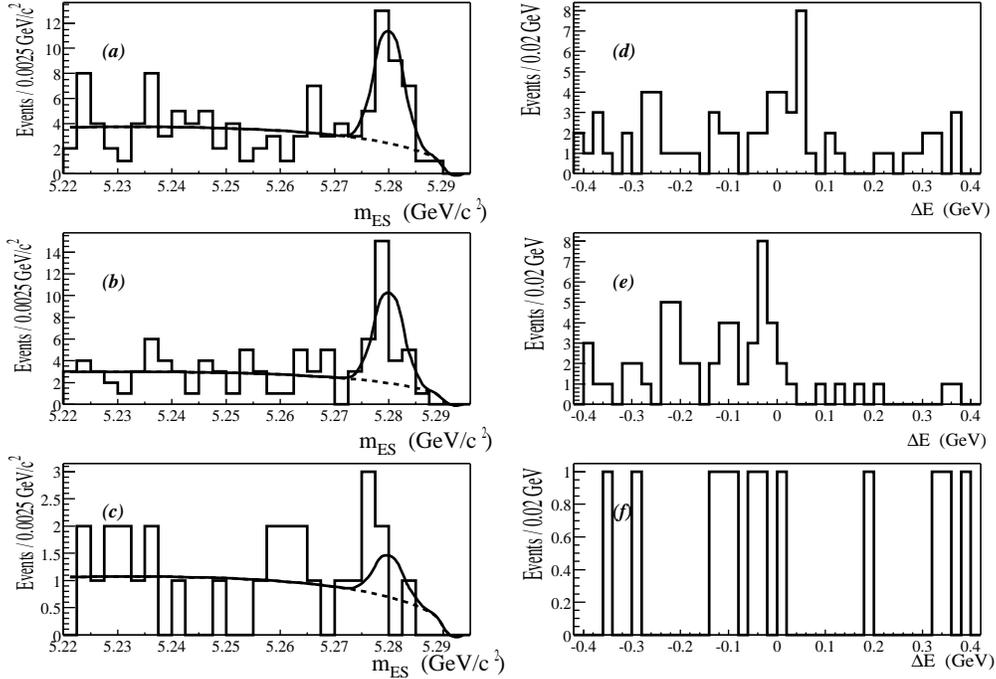}
\caption{Fit results for $\mes$ in the (a) $\pipi$, (b) $\kpi$, and (c) $\kk$ 
samples defined by the particle identification selector described in 
Sec.~\ref{sec:pid}.  Also shown are the $\Delta E$ distributions for the
(d) $\pipi$, (e) $\kpi$, and (f) $\kk$ samples in the $\mes$ signal region.}
\label{fig:cutresult}
\end{center}
\end{figure}

We now decompose the signal into its constituent components by using the 
particle selector to separate the sample into three subsamples corresponding to 
the different channels.  We perform three separate fits similar to the $h^+h^-$ 
fit.  After correcting for the efficiency and mis-identification probabilities 
in Table~\ref{tab:effmisid}, we find $25\pm 8$ $\pipi$, $26\pm 8$ $\kpi$, and 
$1.2^{+3.8}_{-1.2}$ $\kk$ decays.  In Fig.~\ref{fig:cutresult} we show the fit 
results for $\mes$ and the $\Delta E$ distribution in the $\mes$ signal region 
for all three modes.

To facilitate comparison with the global maximum likelihood fit, we determine 
the non-common systematic errors for the cut-based analysis.  We vary the mean 
and width of the $\mes$ signal and background distributions within their 
estimated uncertainties.  Systematic uncertainty on the particle 
identification method is estimated by comparing with various selector 
definitions, including the probability density functions 
used in the global fit (below).  Uncertainty in the ${\cal F}$ shape is 
estimated by varying the cut and by substituting the shape obtained from the 
$\Bub\to\Dz\pim$ data sample (Fig.~\ref{fig:fisher}).  Table~\ref{tab:cutsys} 
summarizes these uncertainties.

\begin{table}[!htb]
\caption{Systematic errors ($\%$) in the cut-based analysis that are not common
to the global fit analysis.  ``PID'' refers to the method of particle 
identification.}
\begin{center}
\smallskip
\begin{tabular}{cccccccccc} \hline\hline
 Mode & Bkg $\mes$ & $\langle\mes\rangle$ & $\sigma(\mes)$ 
 & ${\cal F}\,(\Dz\pi)$ & ${\cal F}$ cut & PID & Total \\\hline
 \bigskip
 $\pip\pim$ & $0.4$ & $^{+0.4}_{-1.2}$ & $0.8$ &  $1.9$ & $11$ & $9$ & $\pm 14$ \\
 \smallskip
 $\Kp\pim$  & $0.4$ & $^{+1.7}_{-3.0}$ & $0.8$ &  $1.9$ & $8$ & $9$ & $\pm 13$ \\
 \smallskip
 $\Kp\Km$   & $0.5$ & $^{+23}_{-27}$ & $^{+2.7}_{-5.3}$ & $1.9$ & $14$ & $14$ & $^{+31}_{-34}$ \\
 \smallskip
 $h^+h^-$ & $0.5$ & $^{+1.2}_{-2.7}$ & $^{+1.3}_{-1.2}$ & $1.9$ & $6$ & -- & $7$ \\ \hline\hline
\end{tabular}
\end{center}
\label{tab:cutsys}
\end{table}

\subsection{Global fit}
\label{sec:fit}

We perform an unbinned maximum likelihood fit using $\mes$, $\Delta E$, ${\cal F}$,
$\theta_1(p_1)$, and $\theta_2(p_2)$, where $\theta_1$ and $\theta_2$ are the \cerenkov\
angles for each track and $p_1$ and $p_2$ are the momenta.  The likelihood ${\cal L}$ is 
defined as
\begin{equation}
{\cal L} = e^{-N^{\prime}} \prod_{i=1}^N 
{\cal P}_i(\mes,\Delta E,{\cal F}, \theta_1(p_1), \theta_2(p_2)|
N_{\pi\pi}, N_{K\pi}, N_{\pi K}, N_{KK}, N_{\rm bkg}),
\label{like}
\end{equation}
where $N_{\rm bkg}$ is the number of continuum background events and ${\cal P}_i$ is 
the probability for the $i$th candidate assuming the total yield 
\begin{equation}
N^{\prime} = N_{\pi\pi}+N_{K\pi}+N_{\pi K}+N_{KK}+N_{\rm bkg}.
\end{equation}
In the \kpi\ mode we fit separately for the two possible combinations 
($\pi K$ or $K\pi$).  The term $(e^{-N^{\prime}})$ derives from the Poissonian
probability of observing $N$ total events when $N^{\prime}$ are expected.  The
probability for a given candidate is the sum of the signal and background terms
\begin{equation}
{\cal P}_i(\mes,\Delta E,{\cal F}, \theta_1(p_1), \theta_2(p_2)|
N_{\pi\pi}, N_{K\pi}, N_{\pi K}, N_{KK}, N_{\rm bkg}) = \sum_k N_k{\cal P}_i^k,
\end{equation}
where the index $k$ represents the five fit components and ${\cal P}_i^k$ is the product of
probability density functions for $\mes$, $\Delta E$, ${\cal F}$, $\theta_1(p_1)$, and
$\theta_2(p_2)$.

\begin{figure}[!htbp]
\begin{center}
\includegraphics[height=12.5cm]{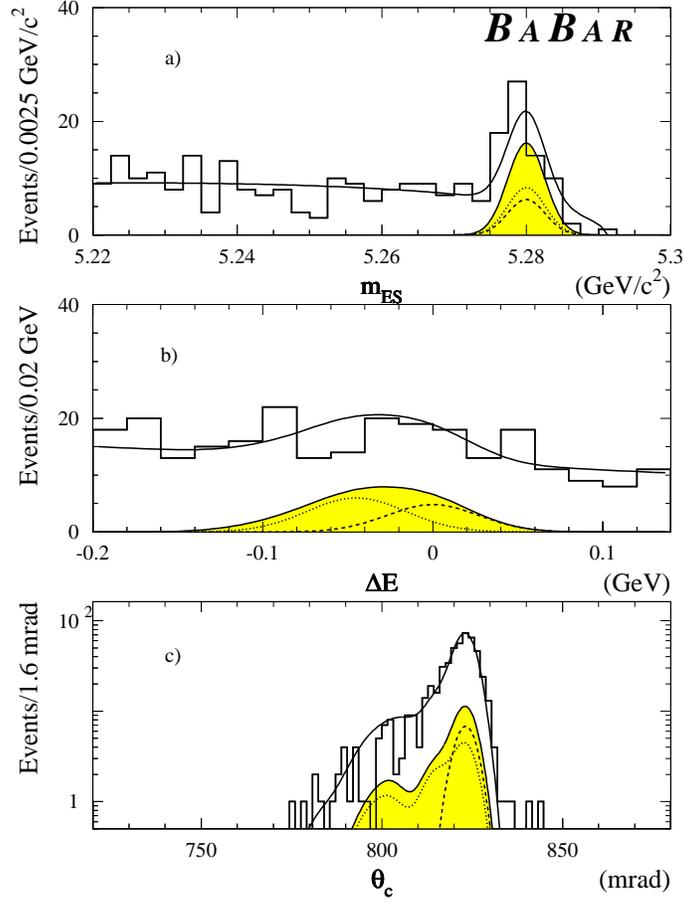}
\caption{The distributions of (a) $\mes$, (b) $\Delta E$, and (c) $\theta_c$ 
(solid histograms) after cutting on $\cos{\theta_S}$, $\cos{\theta_B}$, and
${\cal F}$.  The global fit results are overlaid after scaling by the 
relative efficiency of the additional cuts.  The shaded region is the total signal 
contribution, while the dashed and dotted lines are the $\pipi$ and $\kpi$ 
components, respectively.}
\label{fig:globfit}
\end{center}
\end{figure}

The fit includes all candidates satisfying the selection criteria described in
Secs.~\ref{sec:babar}, \ref{sec:selection}, and \ref{sec:bkg}, as well as the 
requirement on the number of \cerenkov\ photons above background and the proton rejection cut 
described in Sec.~\ref{sec:pid}.  Due to the use of the discriminating variable 
$\Delta E$, the global fit is much less susceptible to contamination from 
three-body \B\ decays.  The signal region is therefore expanded to 
$-0.200 < \Delta E < 0.140\gev$.
The quantity $-\log{{\cal L}}$ is minimized with respect to the fit parameters.  
The resulting signal yields are $29^{+8}_{-7}$ $\pipi$, $38^{+9}_{-8}$ $\kpi$, and 
$7^{+5}_{-4}$ $\kk$ decays.
As a visual cross check, in Fig.~\ref{fig:globfit} we plot the
$\mes$, $\Delta E$, and $\theta_c$ distributions after the additional
requirements on $\cos{\theta_S}$, $\cos{\theta_B}$, and ${\cal F}$ applied in the
cut-based analysis, and overlay the global fit results after rescaling by the
relative efficiency for these cuts.

\begin{table}[!htbp]
\caption{Systematic errors ($\%$) on the signal yield from the global fit that 
are not common to the cut-based analysis.  Uncertainty on $\theta_c$ includes 
mean and width.}
\begin{center}
\smallskip
\begin{tabular}{cccccccccc} \hline\hline
 Mode & Bkg $\mes$ & Bkg $\Delta E$ & $\langle\mes\rangle$ & $\sigma(\mes)$ 
 & $\langle \Delta E \rangle$ & $\sigma (\Delta E)$ & ${\cal F}\,(\Dz\pi)$ & $\theta_c$ & Total \\ \hline 
 \smallskip
 $\pip\pim$ & $0.2$ & $0.4$ & $^{+0.7}_{-2.0}$ & $0.3$ & $+1.9$ & $^{+7}_{-10}$ & $7.0$ & $^{+3.5}_{-2.4}$ & $^{+11}_{-13}$ \\ 
 \smallskip
 $\Kp\pim$  & $0.2$ & $0.2$ & $^{+1.0}_{-5}$ & $^{+0.3}_{-0.4}$ & $-0.7$ & $^{+5}_{-9}$ & $5$ & $^{+1.0}_{-1.4}$ & $^{+7}_{-12}$ \\
 \smallskip
 $\Kp\Km$   & $0.3$ & $6$ & $^{+1.3}_{-3.1}$ & $^{+2.1}_{-2.5}$ & $-3.1$ & $^{+2.7}_{-2.9}$ & $22$ & $2.0$ & $^{+23}_{-24}$ \\\hline\hline
\end{tabular}
\end{center}
\label{tab:globsys}
\end{table}

Systematic errors on the fit results are estimated by varying the signal and
background probability density functions for $\mes$, $\Delta E$, and $\theta_c$ 
within their errors.  The $\Delta E$ width is significantly different between
data and Monte Carlo simulated $\Bub \to \Dz\pim$ decays and there is evidence for a few 
$\mev$ shift of the mean value in the negative direction.  To be conservative 
we vary the width by $\pm 5\mev$ and refit assuming 
$\langle\Delta E\rangle = -5\mev$.  The uncertainty due to the shape of 
${\cal F}$ is determined by using the shape obtained from data $\Bub\to \Dz\pim$ 
decays.  Table~\ref{tab:globsys} summarizes the systematic errors that are not 
common to the cut-based analysis.

\subsection{Comparison and cross-checks}
\label{sec:crosscheck}

In order to compare results for the two methods, we correct the signal yields
obtained in the cut-based analysis by the relative efficiency of the
$\left|\cos{\theta_S}\right|<0.7$, $\left| \cos{\theta_B}\right|<0.8$, and
${\cal F}>0.37$ cuts ($44\%$), and the global likelihood fit yields by the 
efficiency of the \cerenkov\ photon cut and proton rejection cuts ($76\%$).  
In order to compare total yields to the $h^+h^-$ fit result we also correct 
by the common requirement that both tracks have an associated $\theta_c$ 
measurement ($76\%$).  The resulting comparison is summarized in 
Table~\ref{tab:results}.  Given that the two samples are not $100\%$ 
statistically correlated, we find the agreement to be satisfactory.

As a cross-check we perform the global fit after applying the particle selector
and the additional selection criteria used in the cut-based analysis 
(including $\left|\Delta E\right|<0.140\gev$).  The $\Delta E$ and $\theta_c$ 
distributions are used in the fit, the ${\cal F}$ distribution is not.  The 
results, summarized in Table~\ref{tab:check}, indicate that there is no 
significant migration between signal categories.
This result confirms the consistency 
in the use of $\theta_c$ in both analyses.

\begin{table}[!htb]
\caption{Comparison of signal yields corrected for the relative 
efficiency of cut-based and global fit analyses.  The first uncertainty is
statistical, the second is systematic.  For total $h^+h^-$ yield we compare 
the sum of central values for the individual modes in the two analyses with 
the $h^+h^-$ fit result (Fig.~\ref{fig:hh}).}
\begin{center}
\smallskip
\begin{tabular}{cccc} \hline\hline
 Mode & cut-based & global fit & $h^+h^-$ fit \\\hline
 \smallskip
 $\pip\pim$ & $76 \pm 23 \pm 11$ & $50$$^{+14}_{-12}$$^{+6}_{-7}$ & -- \\
 \smallskip
 $\Kp\pim$  & $77 \pm 25 \pm 10$ & $67 ^{+16}_{-14}$$^{+5}_{-8}$ & -- \\
 \smallskip
 $\Kp\Km$   & $4 ^{+11}_{-4}$$\pm 1$ & $13 ^{+9}_{-7}$$\pm 3$ & -- \\
 \smallskip
 $h^+h^-$ & $156$ & $129$ & $151 \pm 24 \pm 11$ \\
 \hline\hline
\end{tabular}
\end{center}
\label{tab:results}
\end{table}

\begin{table}[!htb]
\caption{Signal yields from global fits to the particle-selected samples.  Note
that these results have not been corrected for the $76\%$ efficiency of the
cut on minimum number of \cerenkov\ photons.}
\begin{center}
\smallskip
\begin{tabular}{cccc} \hline\hline
& \multicolumn{3}{c}{Selected as} \\
 Mode & $\pipi$ & $\kpi$ & $\kk$\\\hline
 \smallskip
 $\pipi$ & $21 \pm 5$ & $0.0^{+0.7}_{-0.0}$ & $0.0^{+0.5}_{-0.0}$ \\
 \smallskip
 $\kpi$  & $0.0^{+0.7}_{-0.0}$ & $24 \pm 5$ & $0.0^{+0.7}_{-0.0}$ \\
 \smallskip
 $\kk$   & $0.0^{+0.5}_{-0.0}$ & $1.7 \pm 2.3$ & $2.1 \pm 1.9$ \\
 \hline\hline
\end{tabular}
\end{center}
\label{tab:check}
\end{table}

\begin{table}[!htbp]
\caption{Summary of branching fraction results for the global likelihood
fit.  Shown are the central fit values $N_S$, the statistical significance,
and the measured branching fractions $\BR$.  For the $\kk$ mode, the $90\%$ 
confidence level upper limit on the signal yield is given in parenthesis.  
There is a common efficiency of $0.35\pm 0.02$ for all three modes.  For $N_S$ 
and $\BR$ the first error is statistical and the second is systematic.  Charge 
conjugate modes are assumed.}
\begin{center}
\smallskip
\begin{tabular}{cccc} \hline\hline
 Mode & $N_S$ & Stat. Sig. ($\sigma$) & $\BR\,(10^{-6})$ \\\hline
 \smallskip
 $\pip\pim$ & $29 ^{+8}_{-7}$$^{+3}_{-4}$ & 5.7 & $9.3^{+2.6}_{-2.3}$$^{+1.2}_{-1.4}$ \\
 \smallskip
 $\Kp\pim$  & $38 ^{+9}_{-8}$$^{+3}_{-5}$ & 6.7 & $12.5^{+3.0}_{-2.6}$$^{+1.3}_{-1.7}$ \\
 \smallskip
 $\Kp\Km$   & $7 ^{+5}_{-4}$ ($<15$) & 2.1 & $<6.6$ \\
 \hline\hline
\end{tabular}
\end{center}
\label{tab:brresult}
\end{table}

\section{Determination of branching fractions}
\label{sec:br}

We determine branching fractions for $\pip\pim$ and $\Kp\pim$ decays and 
an upper limit for the $\Kp\Km$ decay using the results of the global 
likelihood fit.  The individual efficiencies were reported in previous 
sections.  The total efficiency is $0.35 \pm 0.02$, where the error is 
combined statistical and systematic.  Branching fractions are calculated as
\begin{equation}
\BR = \frac{N_S}{\epsilon \cdot N_{\BB}},
\label{br}
\end{equation}
where $N_S$ is the central value from the fit, $\epsilon$ is the total efficiency,
and $N_{\BB}$ is the total number of $\BB$ pairs 
in our dataset.  Implicit in Eq.~\ref{br} is the assumption of equal branching 
fractions for $\Y4S\to\Bz\Bzb$ and $\Y4S\to\Bu\Bub$.  The results are 
summarized in Table~\ref{tab:brresult}.  In addition to the systematic 
uncertainties listed in Table~\ref{tab:globsys}, the total error includes uncertainty 
on the tracking efficiency ($2.5\%$ per track) \cite{ref:babar}, the shape of $\cos{\theta_S}$ ($3\%$), 
and the number of $\BB$ events ($3.6\%$).  

\begin{figure}[!htb]
\begin{center}
\includegraphics[height=10cm]{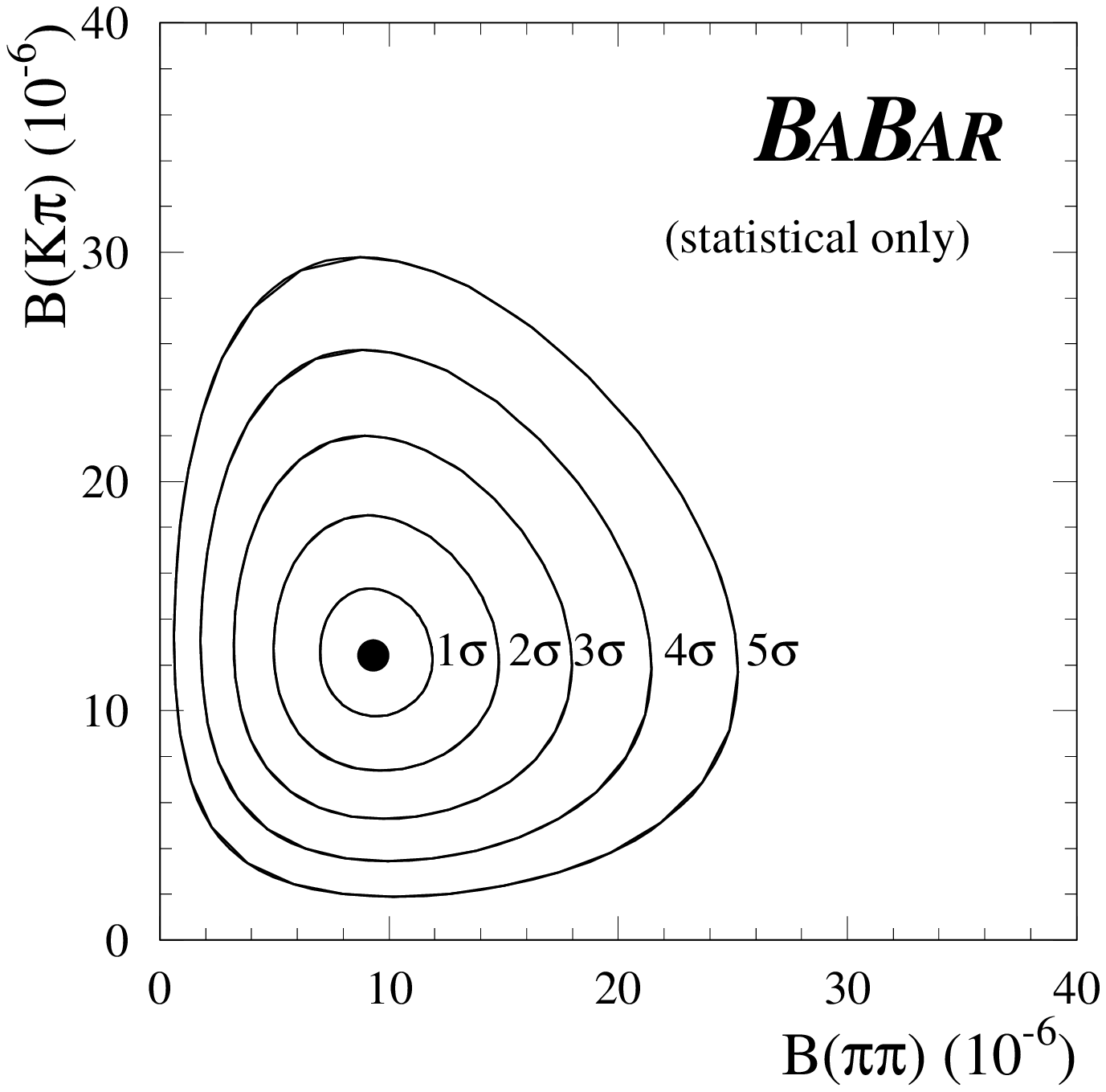}
\caption{The central value (filled circle) for $\BR(\Bz\to \pip\pim)$ and 
$\BR(\Bz\to \Kp\pim)$ along with the $n\sigma$ statistical contour curves for 
the global likelihood fit, where $n\sigma$ corresponds to a change of $n^2$ in 
$-2\log{\cal L}$.}
\label{fig:sig}
\end{center}
\end{figure}

The statistical significance of a given signal yield is determined by setting the yield 
to zero and maximizing the likelihood with respect to the remaining variables. 
The results are give in Table~\ref{tab:brresult}.
Fig.~\ref{fig:sig} shows the $n\sigma$ likelihood contour 
curves, where $\sigma$ represents the statistical uncertainty only.  The curves are 
computed by maximizing the likelihood with respect to the remaining variables in the fit.
For the $\kk$ mode we calculate the $90\%$ confidence level upper limit yield 
and decrease the efficiency by the total systematic error $(24\%)$ before
calculating the upper limit branching fraction.

\section{Summary}
\label{sec:summary}

We have performed a search for charmless two-body \B\ decays to charged 
pions and kaons.  The statistical significance of the signal yields are
$5.7$, $6.7$, and $2.1$ standard deviations for the $\pipi$,
$\kpi$, and $\kk$ decay modes, respectively.  
For the decay modes $\Bz \to \pip\pim$ and $\Bz \to \Kp\pim$
we measure preliminary branching fractions of 
$(9.3^{+2.6}_{-2.3}$$^{+1.2}_{-1.4})\times 10^{-6}$ and 
$(12.5^{+3.0}_{-2.6}$$^{+1.3}_{-1.7})\times 10^{-6}$, respectively, where the 
first uncertainty is statistical and the second is systematic.  
Since the $\kk$ yield is not significant we calculate the $90\%$ confidence 
limit and find $\BR(\Bz \to \Kp\Km) < 6.6 \times 10^{-6}$.

\section*{Acknowledgments}
\label{sec:Acknowledgments}

We are grateful for the contributions of our \pep2\ colleagues in
achieving the excellent luminosity and machine conditions
that have made this work possible.
We acknowledge support from the
Natural Sciences and Engineering Research Council (Canada),
Institute of High Energy Physics (China),
Commissariat \`a l'Energie Atomique and
Institut National de Physique Nucl\'eaire et de Physique des Particules
(France),
Bundesministerium f\"ur Bildung und Forschung
(Germany),
Istituto Nazionale di Fisica Nucleare (Italy),
The Research Council of Norway,
Ministry of Science and Technology of the Russian Federation,
Particle Physics and Astronomy Research Council (United Kingdom), the
Department of Energy (US),
and the National Science Foundation (US). In addition, individual support 
has been received from the Swiss 
National Foundation, the A. P. Sloan Foundation, the Research Corporation,
and the Alexander von Humboldt Foundation.
The visiting groups wish to thank 
SLAC for the support and kind hospitality
extended to them.

\end{document}